\documentclass[]{article}
\usepackage{epsf}
\usepackage{graphicx}

\begin{document}
\baselineskip 21pt

\centerline{\large \bf Structure of the Galaxies in the NGC 80 Group: Two-Tiered Disks}

\centerline{\bf M.A. Startseva(Ilyina)$^1$, O.K. Sil'chenko$^1$ and A.V. Moiseev$^2$}

\bigskip
\bigskip

\noindent
{\it $^1$Sternberg Astronomical Institute of the Moscow State University,
Moscow, Russia}\\

\noindent
{\it $^2$Special Astrophysical Observatory of the Russian Academy
of Sciences, Nizhnji Arkhyz, Russia}

\vspace{2mm}
\sloppypar 
\vspace{2mm}

\bigskip
\bigskip
\bigskip
\bigskip

{\small
\noindent
{\bf Abstract}. $BV$ photometric data obtained at the 6-m telescope of the Special Astrophysical
Observatory are used to analyze the structure of 13 large disk galaxies in the 
NGC 80 group. Nine of the 13 galaxies under considerationare classified by us 
as lenticular galaxies. The stellar populations in the galaxies are very 
different, from old ones with ages of $T>10$ Gyrs (IC 1541) to relatively young
ones with ages of $T<2-3$ Gyr (IC 1548, NGC 85). In one case, current star formation 
is known (UCM 0018+2216). In most of the galaxies, more precisely in all of them 
more luminous than $M_B\sim -18$, two-tiered stellar disks are detected, whose radial 
surface-brightness profiles can be fitted by two exponential segments 
with different scalelengths -- shorter near the center and longer at the periphery. 
All dwarf S0 galaxies with single-scalelength exponential disks are close companions 
to giant galaxies. Except for this fact, no dependence of the properties of S0 galaxies 
on distance from the center of the group is found. Morphological traces of minor merger 
are found in the lenticular galaxy NGC 85. Basing on the last two points, we 
conclude that the most probable mechanisms for the transformation of spirals into 
lenticular galaxies in groups are gravitational ones, namely, minor mergers and 
tidal interactions. 
}

\clearpage

\section{INTRODUCTION}

\noindent
It is wide-spread opinion that lenticular galaxies, which
Hubble \cite{hubble}
thought to be intermediate between elliptical and spiral galaxies, have 
appeared in the past five Gyrs which is confirmed by observations 
of clusters and groups at intermediate redshifts, $z=0.2-0.7$
\cite{fasano00,wilman09}. However, very different hypotheses exist about
details and specific mechanisms for the transformation of spirals into lenticulars. 
It is clear that to transform a spiral into a lenticular galaxy, star formation 
must cease and the stellar disk must be dynamically heated, to eliminate  
spiral arms. However, it is not obvious whether it is enough only 
to cease star formation in the disk or whether it is necessary to increase
strongly (and briefly) the intensity and efficiency of star formation at the
galactic center to facilitate the growth of the bulge, which is, 
on average, more prominent in lenticulars than in spirals of the same mass 
\cite{burstein,mh01}. There is no observational answer to this question up
to date, but we need it to identify the mechanism leading to the transformation 
of spirals into lenticulars. Among gasdynamical mechanisms, the  
ram-pressure of the hot intergalactic medium would predominantly sweep 
out gas of the galaxy, starving (and stopping) star formation \cite{quilis},
while the static pressure of the surrounding hot gas will compress 
the cool gas of the galactic disk, stimulating star formation \cite{zasov}.
As a rule, gravitational effects, such as tidal perturbation of the disk 
dynamics by the potential of the cluster (group) as a whole or by pair 
interactions between galaxies, result in formation of a central bar, 
with the bar then provoking the radial inflow of the outer-disk gas toward 
the galactic center, feeding star formation there. The outer parts of stellar 
disks that are devoid of gas in such way are dynamically heated, and can no 
longer support spiral structure \cite{byrdvalt,moore96,moore99}. Tidal
effects also remove gas from the peripheries of galactic disks. 

Just as the dominant mechanism for transformation of spirals into lenticulars 
is unclear, it is likewise not clear where this transformation occurs. 
At $z=0$, lenticular galaxies are dominant in clusters, where 
they contribute up to 60\%\ of all galaxy population \cite{oemler,dressler}.
This implies that the transformation itself may occur in clusters \cite{fasano00,
ButchOem}, especially because gasdynamical conditions (a dense, hot 
intergalactic medium, a high galaxy velocity dispersion) are favorable for this. 
However, an idea that lenticular galaxies can mainly be formed in groups 
has attracted more and more attention in recent years; within this paradigm, 
groups falling into clusters bring with them lenticular galaxies already completely
formed \cite{wilman09}. This hypothesis is inspired by the observed morphological mix
of galaxies in various types of environments at $z=0.3-0.7$. If it
is correct, it immediately distinguishes gravitational pair interactions, 
especially minor mergers, as the most likely mechanism for the formation 
of S0 galaxies, since groups display lower galaxy velocity dispersions 
with respect to clusters, that weaken gasdynamical effects while enhancing
the effectiveness of gravitational effects. The properties of galaxies in 
groups are currently known not as good as the properties of galaxies in 
clusters which makes studies of the structure and stellar populations of 
lenticular galaxies in groups of key importance for further progress in
understanding galaxy formation. 

Here we study the structure of lenticular galaxies, as well as early-type spiral 
galaxies, which are the most likely candidates for precursors of S0 galaxies, 
all the members of the massive X-ray group of NGC 80. This group is also known 
as GH3 \cite{gh_cat}, SRGb063 \cite{mahdavi00}, and U013 \cite{ramella_cat},
and is sometimes classified as a poor cluster (WBL 009) \cite{wbl99}. The
X-ray flux from the NGC 80 group is
$\lg L_X(h^{-2}_{100}, \mbox{erg/s})=42.56 \pm 0.09$
\cite{mahdavi00}, which is typical for rich groups, and its total mass estimated
from the X-ray luminosity exceeds $10^{14}\, M_{\odot}$ \cite{ramella_cat}.
The group extends over a degree (more than 1 Mpc) on the sky plane. Redshift 
measurements were used to select 45 group members in \cite{mahdavi04}; our
visual inspection of the images indicates that at least half of these 
(24 galaxies) are late-type spiral galaxies, which is unexpected, given 
the powerful X-ray halo of the group. Could it be that the group of NGC 80 
is a young group, so that the process of transforming spiral members into 
lenticular ones has not yet finished? To investigate the properties of a sample 
of lenticular galaxies in this group, we have obtained CCD-images 
in the $B$ and $V$ bands for six-arcminutes fields around five large
galaxies studied by us earlier in \cite{we08}, and have carried out
surface photometry of the targets with dimensions exceeding $10^{\prime \prime}$.
A list of galaxies studied is given in Table 1 together with their global 
(literature) characteristics.

\begin{table*}
\scriptsize{\tiny}
\caption{Global parameters of the galaxies studied}
\begin{flushleft}
\begin{tabular}{clcccccc}
\hline\noalign{\smallskip}
Number$^1$ & Other name & Type (NED$^2$) 
& $D_{25},\, ^{\prime}$ (LEDA$^3$) & $M_B$ (LEDA) 
& $(B-V)_{3"} ^4$  & $V_r$, km/s (NED) & $\Delta$, Mpc$^1$ \\
\hline\noalign{\smallskip}
--  &  NGC~80 & SA$0-$ & 1.82 & --21.6 & 1.03 & 5698 & $\sim 0$ \\
029 &   --  &  -- & 0.36 & --17.6 & 0.84 & 5709 & 0.117 \\
030 &  -- & -- & 0.33 & --17.9 & 0.95 & 6035 & 0.132 \\
034 &  NGC~81 & -- & 0.41 & --18.8 & 0.99 & 6130 & 0.097 \\
035 &  -- & -- & 0.33$^4$ & --18$^4$ & 0.96 & 5372 & 0.111 \\
040 &  NGC~85 & S0 & 0.68 & --19.5 & 1.18 & 6204 & 0.115 \\
016 &  IC~1541 & S0 & 0.72 & --19.4 & 1.04 & 5926 & 0.544 \\
055 & IC~1548 & S0 & 0.68 & --19.3 & 1.00 & 5746 & 0.400 \\
058 &  NGC~93 & S (Sb$^3$) & 1.32 & --20.8 & 1.06 & 5380 & 0.096 \\
041 &  NGC~86 & Sbc (S0/a$^3$) & 0.74 & --20.0 & 1.05 & 5591 & 0.153 \\
044 &  MCG~$+04-02-010$ & S (Sbc$^3$) & 0.89 & --20.3 & 1.17 & 6630 & 0.185 \\
051 & CGCG~479-014B & Sc & 0.63 & --18.5 & 0.93 & 5850 & 0.377 \\
-- &  UCM~0018+2216 & Sb & 0.29 & --17.3 & 0.72 & 5066 & 0.14 \\
\hline
\multicolumn{8}{l}{$^1$\rule{0pt}{11pt}\footnotesize
Mahdavi \& Geller (2004); the distance of 77 Mpc is taken from this paper as well}\\
\multicolumn{8}{l}{$^2$\rule{0pt}{11pt}\footnotesize
NASA/IPAC Extragalactic Database}\\
\multicolumn{8}{l}{$^3$\rule{0pt}{11pt}\footnotesize
Lyon-Meudon Extragalactic Database}\\
\multicolumn{8}{l}{$^4$\rule{0pt}{11pt}\footnotesize
This work}\\
\end{tabular}
\end{flushleft}
\end{table*}

\bigskip
\section{OBSERVATIONS}
\medskip

\noindent
The photometric data were obtained at the 6-m telescope of the Special 
Astrophysical Observatory of the Russian Academy of Sciences with 
the SCORPIO \cite{scorpio} in a direct-image mode. The detector was a
$2048 \times 2048$ EEV 42-40 CCD. The exposures have been acquired
under double binning that provided a scale of $0.35^{\prime \prime}$ per
pixel. The field of view was $6.1^{\prime}$. The observations have been 
carried out in the standard Johnson $B$ and $V$ bands. An exposure of
twilight sky was used as a flat field. 

A detailed list of observations is given in Table 2. The observations of the 
group of NGC 80 were undertaken on August 21, 2007, under photometric 
conditions, and the seeing was about $2^{\prime \prime}$. We took 
exposures of five fields centered onto the most luminous galaxies of the group,
NGC 80, NGC 86, NGC 93, IC 1548, and IC 1541. 
The central galaxy NGC 80 was used as a photometric standard:
a good collection of aperture photoelectric photometry is available for 
this galaxy in the HYPERLEDA database, mainly from \cite{poulain}.

{\footnotesize
\begin{table*}
\caption{Photometric and spectral observations of the group galaxies in 2007-2008. }
\begin{flushleft}
\begin{tabular}{rccccc}
\hline\noalign{\smallskip}
NGC/IC & Observation date & Instrument/mode & Spectral range & Exposure time, s
& Seeing \\
\hline
80 & 21.08.2007 & SCORPIO/IMAGER & $V$ & 60 & $2.3^{\prime \prime}$ \\
80 & 21.08.2007 & SCORPIO/IMAGER & $B$ & 180 & $2.1^{\prime \prime}$ \\
93 & 21.08.2007 & SCORPIO/IMAGER & $B$ & $180 \times 2$ & $2.0^{\prime \prime}$ \\
93 & 21.08.2007 & SCORPIO/IMAGER & $V$ & $60 \times 3$ & $2.0^{\prime \prime}$ \\
86 & 21.08.2007 & SCORPIO/IMAGER & $V$ & $120 \times 2$ & $2.2^{\prime \prime}$ \\
86 & 21.08.2007 & SCORPIO/IMAGER & $B$ & $180 \times 3$ & $2.0^{\prime \prime}$ \\
1548 & 21.08.2007 & SCORPIO/IMAGER & $V$ & $120 \times 2$ & $1.9^{\prime \prime}$ \\
1548 & 21.08.2007 & SCORPIO/IMAGER & $B$ & $180 \times 3$ & $2.1^{\prime \prime}$ \\
1541 & 21.08.2007 & SCORPIO/IMAGER & $B$ & $180 \times 3$ & $1.8^{\prime \prime}$ \\
1541 & 21.08.2007 & SCORPIO/IMAGER & $V$ & $90 \times 3$ & $2.3^{\prime \prime}$ \\
 85 & 04.09.2008 & MPFS & 4200--5600~\AA\ & 5400 & $1.9^{\prime \prime}$ \\
1541 & 02.09.2008 & SCORPIO/Long-Slit & 5700--7000~\AA\ & 1800 & $2.3^{\prime \prime}$ \\
\hline
\end{tabular}
\end{flushleft}
\end{table*}
}

In addition to the photometric data, we have analyzed some spectral observations. 
The central part of NGC 85 was observed with the MPFS (Multi-Pupil Fiber 
Spectrograph) integral-field unit of the 6-m telescope in September 2008
in the blue-green spectral range, 4150--5650~\AA, with a reciprocal dispersion
of 0.75~\AA\ per pixel (and a spectral resolution of about 3~\AA) (for a 
description of this instrument, see \cite{mpfs}). The detector for these
observations was the same $2048 \times 2048$ CCD array. In the MPFS,
a $16 \times 16$ microlens array provides a pupil set, which are connected
to the entrance of the long-slit spectrograph. Such configuration enables 
the simultaneous registration of 256 spectra, each of which corresponds 
to a spatial element of the galaxy image 
$1^{\prime \prime} \times 1^{\prime \prime}$ in size; accordingly, the 
MPFS field of view is $16^{\prime \prime} \times 16^{\prime \prime}$. 
The MPFS data were used to investigate the rotation of the stellar 
component in the center of NGC~85 (we were not able to detect emission
lines in the galaxy), through a cross correlation with the spectra 
of K giant stars observed on the same night with the same spectrograph. 
The age and metallicity of the stellar population in the nucleus and 
circumnuclear region have been estimated by calculating the Lick indices
H$\beta$, Mgb, Fe5270, and Fe5335 \cite{woretal94} and by comparing them
to the models for "simple stellar populations" (SSP) by
Thomas et al. \cite{tmb03}.

To study the kinematics at distances of more than $8^{\prime \prime}$ 
from the center, the lenticular galaxy IC~1541 which is located at 
the periphery of the group was observed on September 2, 2008 with the 
SCORPIO reducer in a long-slit mode. The slit orientation was close to  
the major axis of the galaxy. The exposure was 30 min, and the red grism 
was used as a disperser, providing a spectral range of 5700--7400~\AA\ and
a spectral resolution of about 5~\AA. The original aim was to search for
weak emission lines, H$\alpha$ or [NII]$\lambda$6583, from the ionized gas 
in IC 1541. However, no traces of emission were detected, and 
the data were instead used to estimate stellar line-of-sight velocities 
along the major axis (a rotation curve) via cross correlation of the galaxy 
spectra at various distances from the center with the spectra of twilight sky 
taken on the same night with the same instrumentation.

\bigskip
\section{ANALYSIS OF THE STRUCTURE FOR THE DISK GALAXIES IN THE NGC 80 GROUP}

\medskip
\begin{figure*}
\includegraphics[width=\hsize]{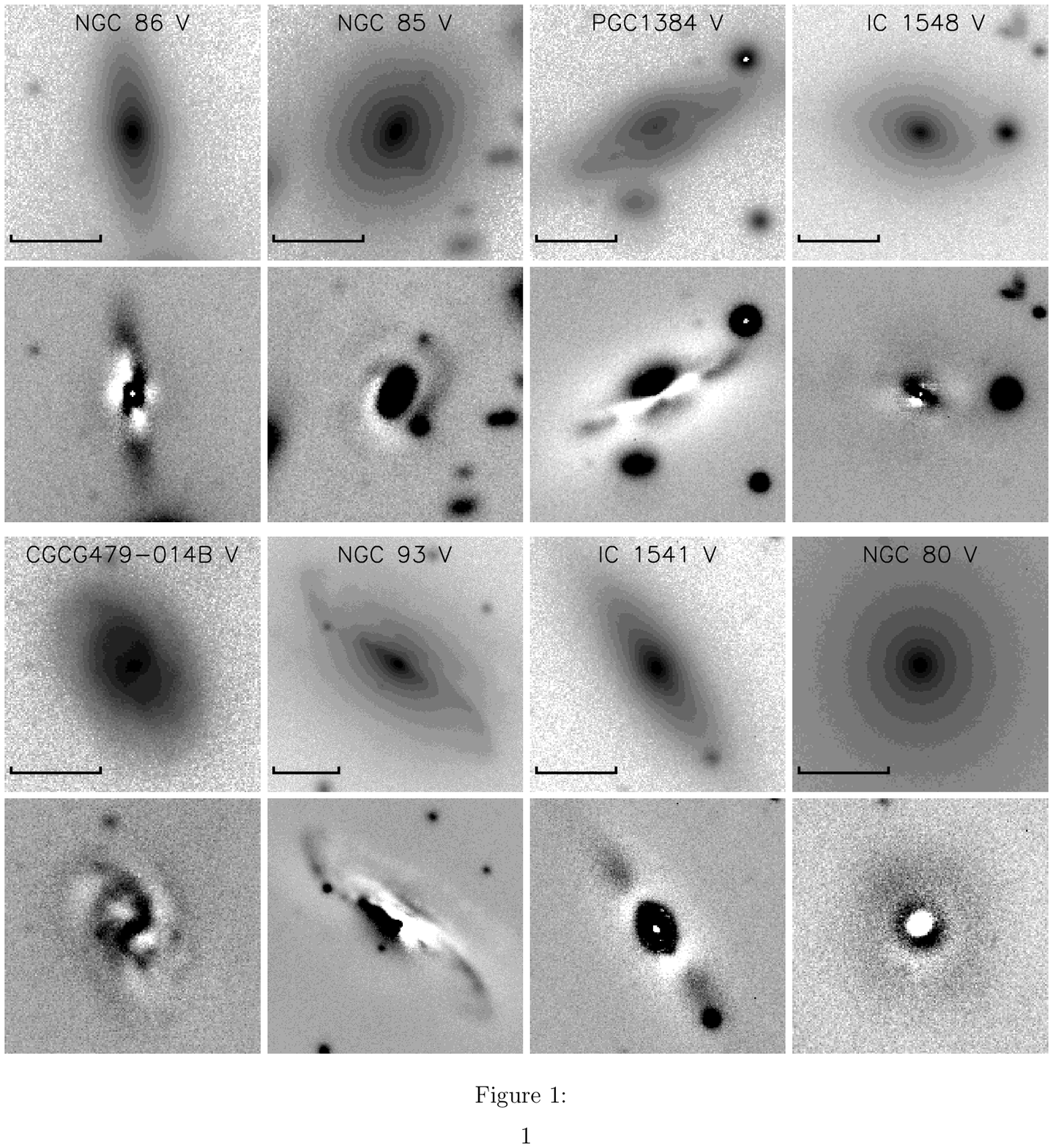}
\caption{$V$-images obtained with the SCORPIO for eight 
luminous group members (the first and the third horizontal rows show initial 
images; the intensity scale is logarithmic), and below every image there are 
maps of the residual brightness after subtracting multi-tier models for the disks 
(the second and fourth horizontal rows; the intensity scale is linear). The bars 
indicate the $20^{\prime \prime}$ length.}
\end{figure*} 

\begin{figure*}
\includegraphics[width=\hsize]{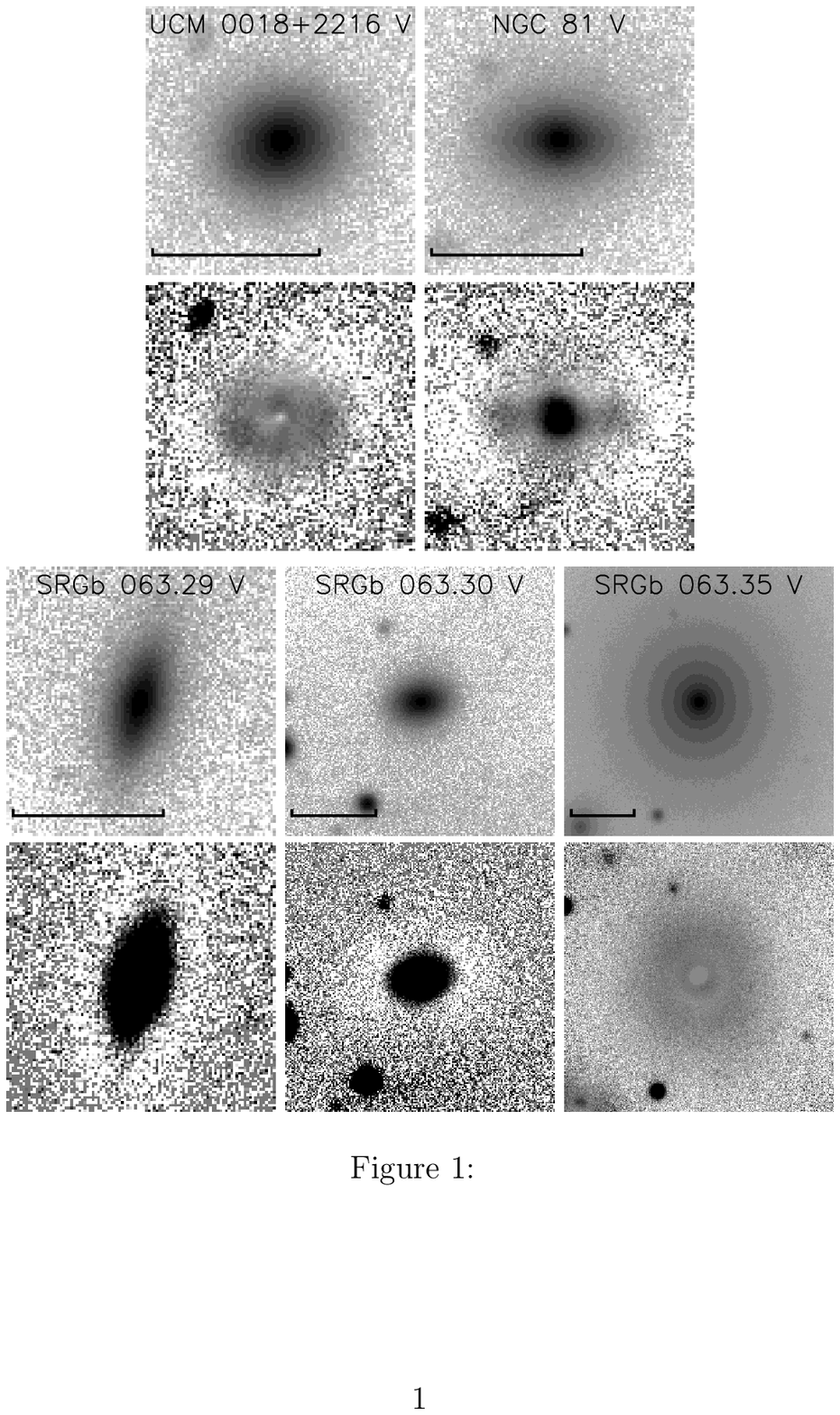}
\caption{The same as Fig. 1 but for five dwarf group members. The residual
brightnesses are given in logarithmic intensity scale.}
\end{figure*} 

\noindent
We have carried out decomposition of the $B$ and $V$ radial
surface brightness profiles of the galaxies under consideration.
Figures 1 and 2 show the original $V$ images of the galaxies,
together with the residual brightnesses after subtracting the models. 

\begin{figure*}
\begin{tabular}{cc}
\includegraphics[width=0.45\hsize]{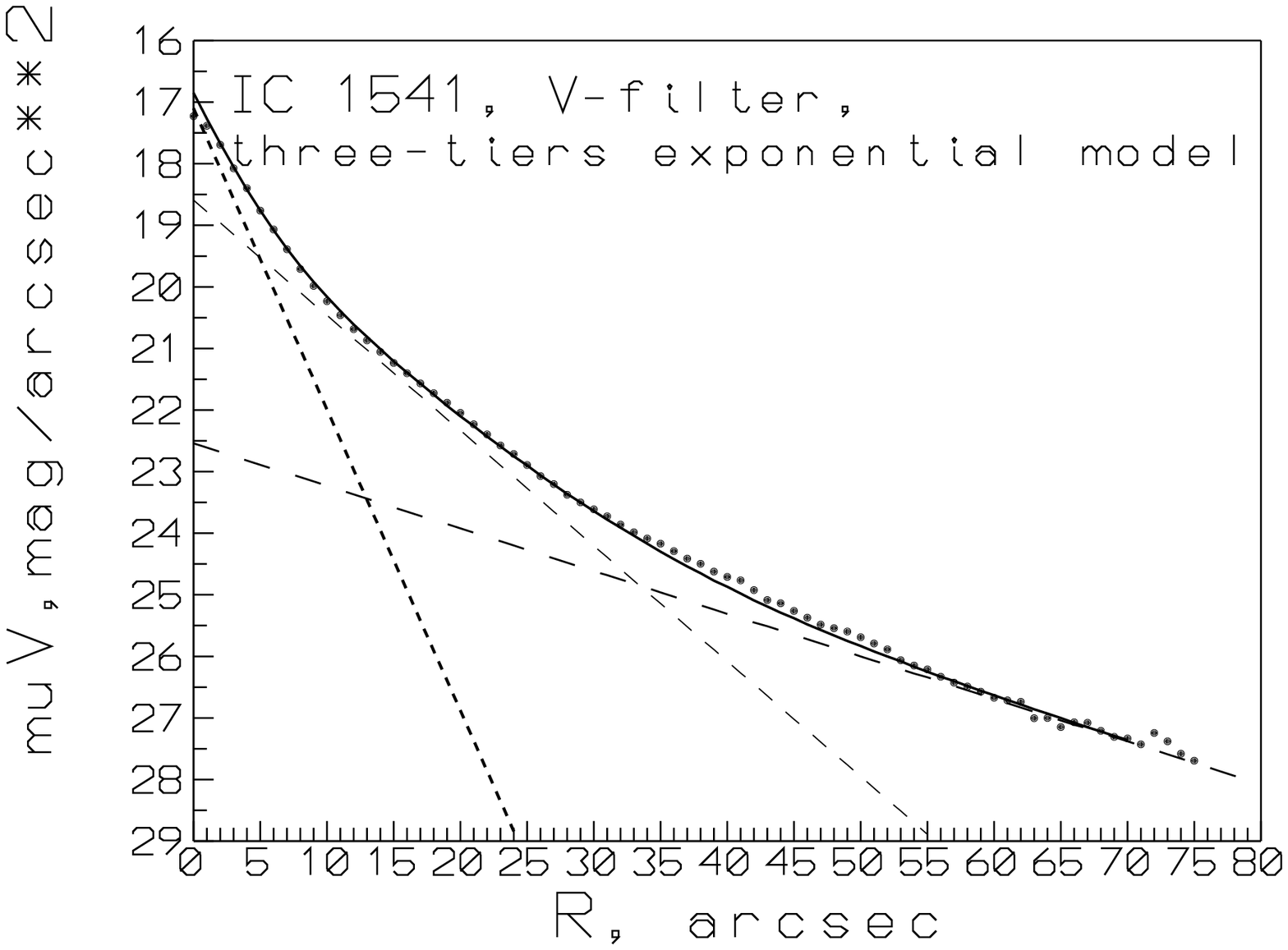} &
\includegraphics[width=0.45\hsize]{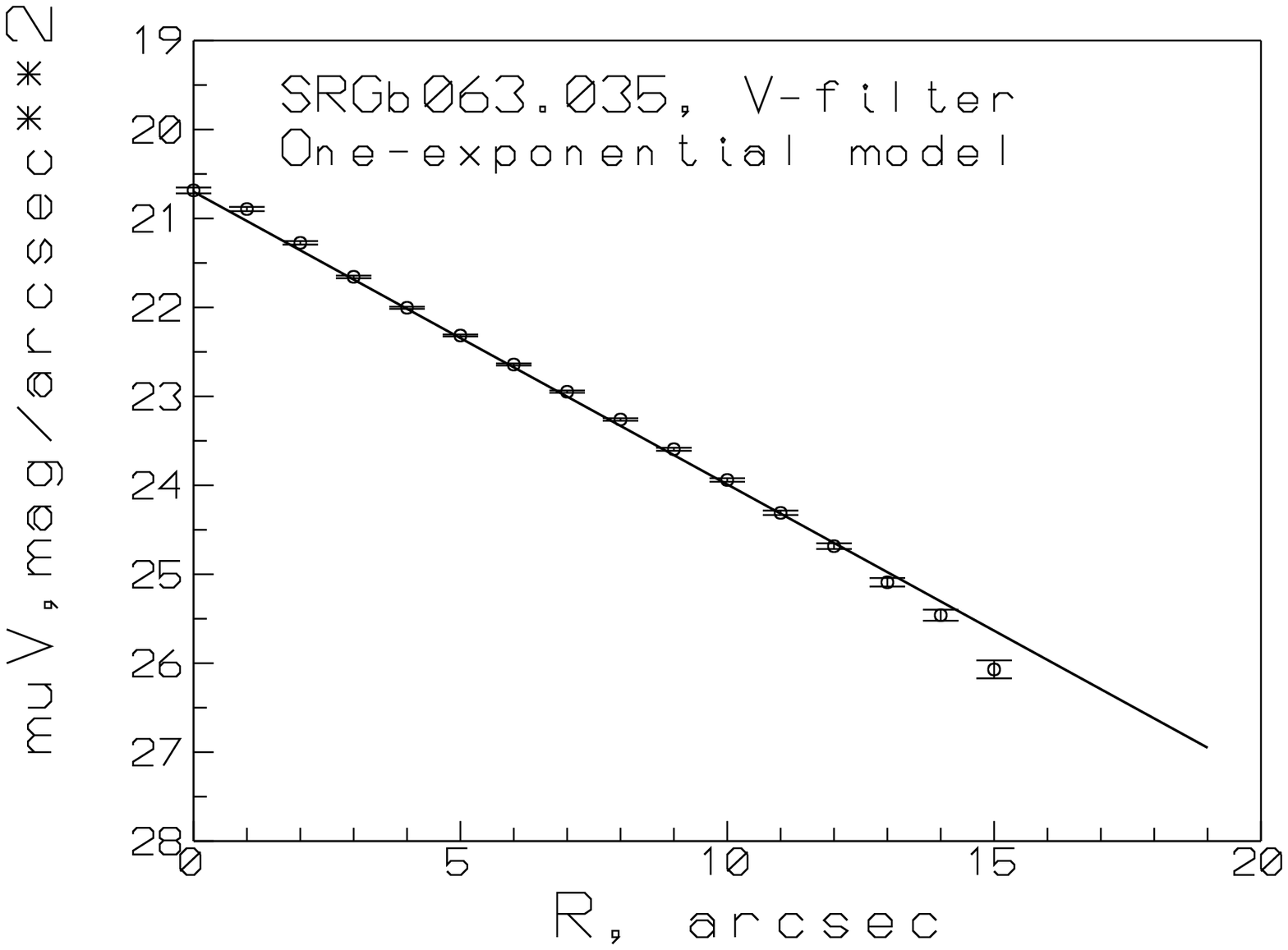}
\end{tabular}
\caption{Examples of decomposing the surface brightness profiles into components.
{\it a} -- The giant lenticular galaxy IC~1541. The lond-dashed line is an
outer disk model, the short-dashed line is an inner disk model, and the dotted
line shows the model of a pseudobulge. The points represent the observational
$V$-brightness profile azimuthally averaged along the ellipses with parameters
derived from the isophotal analysis. The solid curve gives the final fit by
a sum of model components. {\it b} -- The dwarf galaxy SRGb 063.035 -- the
nearest companion to NGC 80. The total profile is perfectly fitted by a single
exponential law.}
\end{figure*}

The decomposition was carried out by using the interactive GIDRA program, 
which successively constructs one-dimensional profiles and two-dimensional 
models (for the description of its application see \cite{moiseev}).
The whole procedure starts from constructing radial profiles of 
the surface brightness averaged in elliptical rings. The shape of these 
elliptical rings (position angles, major axes, and axial ratios)
is specified based on isophotal analysis of the images made with the 
program FITELL from V.V. Vlasyuk. Images of foreground stars projected 
against the galaxies were either masked or removed by subtracting a 
symmetrical profile corresponding to the scattering function for a point source 
(the case of a bright star next to NGC 86). Further, having constructed
the azimuthally averaged surface-brightness profiles, the GIDRA program 
fits it with a Sersic law \cite{sersic}:

$$
\mu (r)= \mu _e+ 1.086 b_n[(r/r_e)^{\frac{1}{n}} - 1],
$$
\noindent
where $n$, $r_e$, and $\mu _e$ are parameters of the model and
the coefficient $b_n \approx 2n - 0.32$. We separate structural 
components of a galaxy, starting with the outer regions which we assume 
to be exponential disks \cite{freeman} -- a particular case of a
Sersic law with $n=1$. Having fitted an exponential to the brightness
profile of the outer stellar disk and determined its parameters, we 
construct a two-dimensional model (image) of this disk and subtract it 
from the observed galaxy image. After that, the procedure of constructing 
and fitting an azimuthally averaged surface brightness profile is repeated
once more. If the ellipticity of the residual isophotes is not very different 
from the ellipticity of the outer isophotes, we conclude that a separate 
inner stellar disk is present in the galaxy, and fit it using an exponential law
with new parameters. After subtracting the inner disk, there usually remained 
a central bulge, for which the Sersic parameter $n$ was set free. 
However, in the most cases, we obtained an exponential brightness profile for 
the bulges as well. This is the first conclusion of our analysis of the 
structure of the early-type disk galaxies in the NGC 80 group: most 
of them have exponential bulges and two-tiered exponential disks. An 
example of decomposing the brightness profile for a giant lenticular 
galaxy (IC 1541 at the periphery of the group) into two disks and an 
exponential bulge is shown in Fig. 3a. This galaxy is seen nearly edge-on, 
and a boxy central structure associated with the bulge is visible 
in the the galaxy; outside the bulge, the two stellar disks with different 
exponential scales have comparable thicknesses. Figure 3a demonstrates how 
deep our photometry is: the galactic disk extends to a radius of 30 kpc, and 
the surface brightness limit is $\mu _V = 27.7^m /arcsec^2$ being well beyond
the Holmberg radius. 

\begin{table*}
\scriptsize{\tiny}
\caption{Parameters of the photometric components of
the galaxies in the group NGC 80}
\begin{flushleft}
\begin{tabular}{|r|c|r|c|ccl|ccl|cccl|}
\hline
Galaxy & Band & $PA^1$ & $(1-b/a)^1$ &
\multicolumn{3}{|c|}{Outer disk} &
\multicolumn{3}{|c|}{Inner disk} & \multicolumn{4}{|c|}{Bulge}\\
&  & & & $\mu _0$, $^m$/arcsec$^2$ &
$h^{\prime \prime}$ &
$h$, kpc &
$\mu _0$, $^m$/arcsec$^2$ &
$h^{\prime \prime}$ &
$h$, kpc &
$n$ &
$\mu _0$, $^m$/arcsec$^2$ &
$h^{\prime \prime}$ &
$h$, kpc \\
\hline
(NGC)80 & $B$ & $4^{\circ}$ & 0.11
& 22.6 & 25.1 & 9.4 & 20.7 & 5.7 & 2.1 & 1 & 18.4 & 2.0 & 0.7 \\
(NGC)80 & $V$ & &
& 21.4 & 27.7 & 10.3 & 19.4 & 4.8 & 1.8 & 1 & 17.3 & 1.8 & 0.7 \\
(SRG)29 & $B$ & $166^{\circ}$ & 0.52
& 24.1 & 8.8 & 3.3 & -- & -- & -- & -- & -- & -- & -- \\
(SRG)29 & $V$ & &
& 23.1 & 8.7 & 3.2 & -- & -- & -- & -- & -- & -- & -- \\
30 & $B$ & $100^{\circ}$ & 0.26
& 23.2 & 9.5 & 3.5 & -- & -- & -- & -- & -- & -- & -- \\
30 & $V$ & &
& 23.0 & 9.4 & 3.5 & -- & -- & -- & -- & -- & -- & -- \\
34 & $B$ & $110^{\circ}$ & 0.1
& 24.6 & 17.3 & 6.5 & 20.1 & 2.2 & 0.8 & 1 & 19.0 & 1.1 & 0.4 \\
34 & $V$ & &
& 23.3 & 16.7 & 6.2 & 19.1 & 2.2 & 0.8 & 1 & 17.9 & 1.1 & 0.4 \\
35 & $B$ & $159^{\circ}$ & $0.1 - 0.18$
& -- & -- & -- & 22.0 & 4.0 & 1.5 & -- & -- & -- & -- \\
35 & $V$ & &
& -- & -- & -- & 20.7 & 3.3 & 1.2 & -- & -- & -- & -- \\
40 & $B$ & $148^{\circ}$ & 0.18
& 22.6 & 9.2 & 3.4 & 21.6 & 5.7 & 2.1 & 1 & 20.0 & 2.2 & 0.8 \\
40 & $V$ & &
& 21.6 & 9.2 & 3.4 & 20.3 & 6.2 & 2.3 & 1 & 19.0 & 2.4 & 0.9 \\
16 & $B$ & $35^{\circ}$ & 0.64
& 24.5 & 24.6 & 9.2 & 21.3 & 7.5 & 2.8 & 1 & 19.2& 2.1 & 0.8 \\
16 & $V$ & &
& 22.5 & 15.7 & 5.9 & 18.6 & 5.8 & 2.2 & 1 & 17.1 & 2.2 & 0.8 \\
55 & $B$ & $85^{\circ}$ & 0.36
& 23.1 & 11.5 & 4.3 & 21.5 & 6.3 & 2.4 & 1 & 19.5 & 2.2 & 0.8 \\
55 & $V$ & &
& 22.4 & 12.5 & 4.7 & 20.6 & 6.3 & 2.4 & 1 & 18.6 & 2.5 & 1.0 \\
58 & $B$ & $53^{\circ}$ & 0.36
& 23.4 & 26.1 & 9.7 & 20.6 & 7.2 & 2.7 & 2 & 20.6 & 4.1 & 1.5 \\
58 & $V$ & &
& 22.5 & 25.5 & 9.5 & 19.4 & 7.6 & 2.8 & 2 & 19.5 & 3.1 & 1.2 \\
41 & $B$ & $8^{\circ}$ & 0.64
& 22.8 & 9.0 & 3.4 & 19.1 & 4.8 & 1.8 & 1 & 19.3 & 2.0 & 0.8 \\
41 & $V$ & &
& 22.2 & 11.4 & 4.2 & 18.9 & 4.9 & 1.8 & 1 & 18.4 & 2.0 & 0.8 \\
44 & $V$ & $116^{\circ}$ & 0.53
& 23.4 & 23.0 & 8.6 & 21.0 & 9.0 & 3.4 & -- & -- & -- & -- \\
51 & $B$ & $32^{\circ}$ & 0.32
& 22.6 & 9.0 & 3.4 & 20.4 & 6.9 & 2.6 & -- & -- & -- & -- \\
51 & $V$ & &
& 22.7 & 9.6 & 3.6 & 19.5 & 6.9 & 2.6 & 1 & 22.9 & 1.4 & 0.5 \\
UCM & $B$ & $105^{\circ}$ & 0.10
& 25.0 & 7.5 & 2.8 & -- & -- & -- & 1 & 20.3 & 1.8 & 0.7 \\
UCM & $V$ & &
& 24.0 & 7.0 & 2.6 & -- & -- & -- & 1 & 19.5 & 1.8 & 0.7 \\
\hline
\multicolumn{14}{l}{$^1$\rule{0pt}{11pt}\footnotesize
Orientation parameters (major-axis $PA$ and isophote ellipticity)
relate to the outermost isophotes.}\\
\end{tabular}
\end{flushleft}
\end{table*}

The parameters of the photometric models for the sample galaxies are 
presented in Table 3, which gives the central surface brightness 
$\mu _0= \mu (0)$, {\it not} reduced to a face-on orientation. All the
luminous lenticular galaxies have two-tiered exponential disks and exponential 
bulges. The bulge of the giant spiral NGC 93 is well fitted with a Sersic
parameter of two, that is typical for giant Sb galaxies. An interesting feature 
is found in the dwarf lenticular galaxies which are satellites of NGC 80 -- the central 
galaxy of the group: three of four display a single-scaled stellar disk, and, 
when present, the central spheroid is best-fitted with a Sersic law with the
index $n<1$, that is typical for dwarf diffuse spheroidal galaxies. In one case
-- namely in the nearest close companion to NGC 80, SRGb063.035 -- there is 
no bulge at all, and the whole galaxy consists of a single exponential disk (Fig. 3b). 
A structure with very low surface brightness is also visible around the 
single exponential disk of a close companion to NGC 86 -- UCM 0018+2216, 
which is classified as a Sb galaxy according to the NED. However, no spiral 
arms are seen in the disk of UCM 0018+2216 in our plots, both $B$ and $V$, 
and we re-classify it as a S0 galaxy that is probably in the process of forming. 
It is obvious that the tidal influence of massive neighbors truncates the brightness 
profiles of the dwarf galaxies and stimulated their secular evolution, smoothing 
the whole stellar density profile into a single exponential law. This provides 
an interesting hint to possible origins of the exponential shapes of 
surface brightness profiles of stellar disks which are not understood yet. 

\begin{figure}
\includegraphics[width=\hsize]{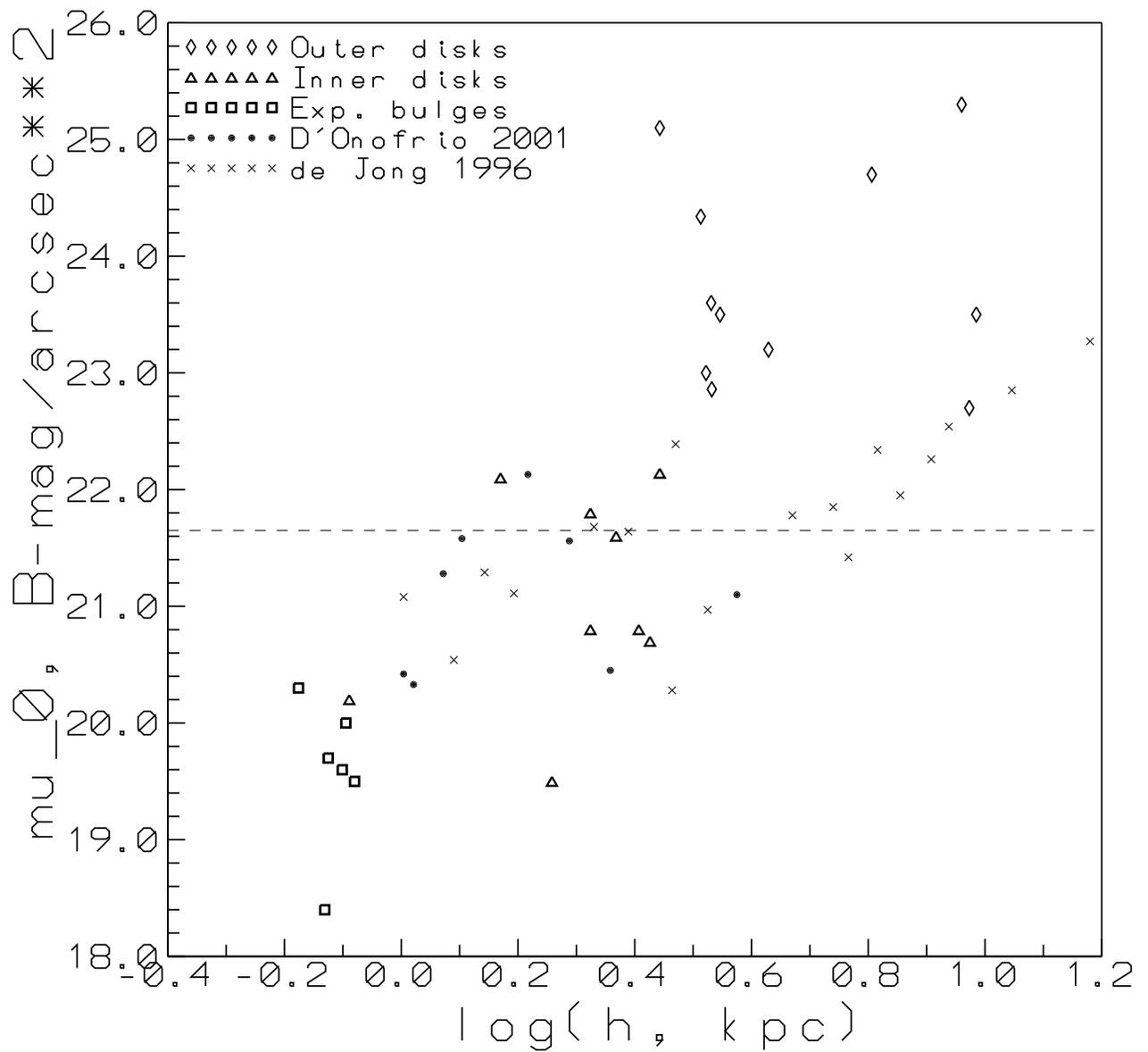}
\caption{Empirical relation between the central surface brightnesses and the
exponential scalelengths for stellar disks of early-type galaxies. The results
of our decomposition -- the outer disks, the inner disks, and the pseudobulges --
are plotted together with the literature data obtained within the framework of
the one-exponential disk model. The horizontal dashed line indicates the mean
central surface brightness of nearby luminous stellar disks by Freeman (1970).}
\end{figure}

Figure 4 compares the central surface brightnesses and scalelengths for 
the exponentials fitted to all three types of structural components --
the outer and inner disks and the exponential bulges. The central 
surface brightnesses reduced to face-on orientation 
are plotted; we necessarily took into account the internal absorption 
in the galaxies, as well as the fact that stellar disks in the 
lenticular galaxies are thick.  Historically, after the publication 
of the first statistics of the stellar disk parameters for nearby 
galaxies by Freeman \cite{freeman}, who reported that the central surface
brightnesses were grouped tightly near the value of $\mu _0(B)=21.7^m /arcsec^2$
and had a Gaussian distribution with the width of the order of the measurement 
errors, all subsequent photometric surveys gave broad distributions for the 
central surface brightnesses and a correlation between the central surface 
brightness and the exponential scalelength (see, for example, 
\cite{dejong,iodice,graham,donofrio}). However, these surveys
did not allow for the possible presence of multi-tiered exponential disks. 
A correlation between the central surface brightness and the 
exponential scalelength is also obvious in our Fig. 4, where 
up to three structural components are plotted for every galaxy. 
However, we can now see that the origin of this correlation 
is due to the presence of three types of structural components with 
a various degree of concentration toward the center. The three clouds 
of points corresponding to the outer disks, inner disks, and exponential 
bulges have different mean scales and central surface brightnesses, but 
no correlation between the scalelengths and the central surface brightnesses 
is observed within each cloud of points. The outer disks all have lower 
surface brightnesses than the "Freeman's standard", while the 
inner disks are all near this line and brighter. We cannot rule out 
that there is no physical correlation between scalelenth and surface 
brightness, but there exist several types of stellar disks of 
different origins. However, this conclusion waits for verification 
with a larger statistics of surface brightness profile decomposition 
taking into account the presence of multi-tiered disks. 

After the construction of the model images for the galaxies studied and the 
subtraction of these from the observed ones, apart from the large-scale 
structural components, we have derived also a lot of small-scale peculiar 
structures which are clearly visible in the residual brightness 
maps in Figs. 1 and 2. Below we present a brief description of these 
substructures for every galaxy. 

{\bf NGC 86}. The stellar population in the galactic center is old \cite{we08}.
After subtracting the outer and inner exponential disks, a boxy bulge with 
spiral arms is left in the residual brightness distribution, with the 
position angle of the major axis of the bulge, about $170^{\circ}$, differing 
substantially from the orientation of the disks. The best fit to the bulge 
is obtained using a Sersic law with the index of one (an exponential bulge) 
and with $r_0=2^{\prime \prime}$, $\mu _0 =19^m /arcsec^2$, and the ratio
of the minor and major axes being about 0.95. Subtracting this bulge leads 
to a residual image that is reminiscent of a bar that is evolved in the 
$z$-direction. The residual brightness after this subtraction occupies
roughly 50\%\ of the total area of the bulge, i.e., it is more compact.
We suggest that the exponential bulge of the galaxy is a pseudobulge whose 
origin is associated with the development of vertical instability 
in a bar \cite{bureau06}.

{\bf NGC 85}. The outer disk starts to dominate the brightness distribution 
at distances $r> 20^{\prime \prime}$, or 7.5 kpc, from the center, and its 
parameters are typical for the disks of spiral galaxies (Table 3). After 
subtracting the outer disk, the residuals have an exponential profile with 
excess brightness near the southern edge. It seems that the remaining inner 
disk has a higher inclination and a smaller position angle than the outer disk. 
The new orientation parameters are $i=50^{\circ}$, $PA =137^{\circ}$. Fitting 
an exponential to the brightness profile in the radius range of 
$8^{\prime \prime} -15^{\prime \prime}$ yields the parameters of a second 
disk (Table 3). After subtracting this disk as well, the residuals resemble 
a small lens-like region with tails. It is interesting that the position 
angle of the small lens seems to be closer to the position angle of the 
outer disk than to that of the inner disk, whereas the opposite is true 
for its inclination. We have recently seen a similar structure in the 
lenticular galaxy with counterrotating gas NGC 5631, which clearly demonstrates 
signatures of minor merger \cite{we09}. Further fitting of the residual
brightness profile for NGC 85 seems to indicate a fairly elongated exponential 
bulge ($b/a=0.83$). The residuals after subtraction of this bulge visually 
resemble three features with tails, which are likely a consequence of a 
recent merger, which also led to the formation of the lenticular galaxy. 

{\bf PGC 1384}. It is immediately obvious that this galaxy contains large amount
of dust. Since the galaxy is viewed nearly edge-on, the dust strongly impedes 
the decomposition, as the galactic center is hidden behind this dust. It is 
interesting that the dust is not visible in the outer disk of this very dusty 
galaxy, which apparently testifies to secular evolution during which the gas 
and dust have become concentrated near the galactic center. Therefore, we adopted 
as the model center the center of the outer isophotes. The fitting of the outer 
disk was carried out for $r>30^{\prime \prime}$. At $r<19^{\prime \prime}$ 
the residual profile in the $V$-band is well-fitted by an exponential law
corresponding to a second disk. We are not able to analyze the most central 
region of the galaxy, since it is strongly contaminated by the dust, 
even in the $V$-band.

{\bf UCM 0018+2216}. The name of this galaxy means that it has been detected 
in the Universidad Complutense de Madrid survey, which is aimed to search 
for emission-line galaxies. Subsequent long-slit spectroscopy \cite{ucm96}
has shown that star formation of modest intensity is occurring in the galactic
nucleus. Our images display a weak asymmetry of the galaxy relative to the 
center of the inner isophotes: its south-eastern half is more elongated
than its north-western part. Because of this, the northern parts of the 
galaxy are slightly oversubtracted, and the residual brightness distribution
is shifted to the south of the center. The outer disk dominates at 
$r>13^{\prime \prime}$. After subtracting the outer disk, a second 
exponential disk can be fit to the residual profile at 
$r= 2^{\prime \prime} - 6^{\prime \prime}$. After subtracting both disks,
there remains an interesting region resembling a boxy bulge with an 
elongated shape and a kind of low-contrast, tightly wound spiral structure 
(a ring?) visible in the $B$-band. 

{\bf IC 1548}. The stellar population in the galactic center is young,
younger than 1.5 Gyr in the nucleus and about 3 Gyr in the bulge \cite{we08}.
After subtracting the two exponential disks from this fairly strongly inclined 
lenticular galaxy, there remains a circular exponential bulge surrounded by 
rudimentary spirals. 

{\bf CGCG 479-014B}. This is a late-type spiral with a large bar. 
The brightness profile of the outer disk is of type II according to 
the definition by Freeman \cite{freeman} -- exponential with a hole
in the center. The bulge is small and circular, and its Sersic parameter 
is determined only uncertainly due to its compactness. 

{\bf NGC 93}. Our previous study \cite{we08} has revealed a chemically
distinct nucleus inside the old bulge of this galaxy, which has formed 3–-4 
Gyr ago in a secondary burst of star formation. The powerful dust lane 
visible in images indicates that the north-western part of the 
disk is the nearest to us. When constructing our models, we had to mask 
both the region of strong dust absorption and the high-contrast 
spiral arms. The outer disk begins to dominate at $r=43^{\prime \prime}$ 
(16 kpc), while the inner disk dominates at 
$r=20^{\prime \prime} - 40^{\prime \prime}$. After subtracting the two 
disks, there remains a small lens-like region that is asymmetrical 
due to crossing by the dust lane. It is possible to fit the residual 
brightness at $r=5^{\prime \prime} - 11^{\prime \prime}$ by
a single ring-like disk (with a hole in its center), with its brightness 
profile cut off at $r>11^{\prime \prime}$. After fitting the residual 
brightness at $r=2^{\prime \prime} -11^{\prime \prime}$ with a Sersic
bulge with $n=2$ and isophote ellipticity of 0.75, a very
interesting residual brightness distribution appears, which resembles 
a one-sided bar (whose other half is hidden by the dust), or a satellite 
that is disrupted by tidal forces. 

{\bf IC 1541}. The stellar population at the galactic 
center is old \cite{we08}. See Fig. 3a for an analysis of the
surface-brightness profile. The apparent axial ratio for 
the exponential bulge is 0.8. 

{\bf NGC 80}. The decomposition results for this galaxy agree well, as 
concerning the scalelengths, with the results published by us 
previously based on less deep data \cite{we03}. The outer regions
can be fit with an exponential profile from a radius of $30^{\prime \prime}$ 
(11 kpc). The inner disk dominates at radii of $10^{\prime \prime}-
20^{\prime \prime}$ (4--7 kpc). After subtracting the two disks, the residual
brightness at $r=2^{\prime \prime} -8^{\prime \prime}$ can be 
fit with a circular exponential bulge. Excess brightness is visible 
at $r=5^{\prime \prime} -7^{\prime \prime}$, consistent with the position
of a ring of relatively young stars ($T\sim 5$ Gyr) \cite{we03,we08}.

{\bf NGC 81}. An outer exponential disk fits the galaxy profile well 
at $r>15^{\prime \prime}$, and an inner disk -- from $r>6^{\prime \prime}$ 
to $12^{\prime \prime}$. After subtracting these two disks, the residual 
structure resembles a lens with weaks spiral arms, which we fit by an 
exponential bulge with axial ratio of $b/a=0.95$. After subtracting the bulge, 
there remain residual tails resembling a spiral structure. The bulge 
has boxy isophotes. 

{\bf SRGb 063.030}. The outer disk reveals an exponential profile starting from 
$r=11^{\prime \prime}$ (4 kpc). After disk subtraction, the residual profile has 
a bell-like shape that cannot be fitted by a Sersic law with $n \ge 1$.

{\bf SRGb 063.029}. The outer disk starts at $r=15^{\prime \prime}$ 
(5.5 kpc). After subtracting this outer disk, the residual profile has a 
bell-like shape. 

{\bf SRGb 063.035}. This galaxy is very compact, and can be fitted by a 
single exponential disk (Fig. 3b). It is most likely that its outer regions
have been truncated by the tidal influence of NGC 80, which is at a distance 
of only 20 kpc in projection onto the sky plane.

\bigskip
\section{ROTATION CURVE OF THE LENTICULAR GALAXY IC 1541}
\medskip

\noindent
We have obtained the long-slit spectrum of IC 1541 with the SCORPIO,
the slit being aligned roughly with the major axis of the 
galaxy. We aimed to search for ionized gas, but no signs 
of emission lines were detected in the long-slit spectrum over 
the full extent of the galactic disk. However, we have
used this spectrum to estimate the rotational velocity 
of the stellar component: we were lucky to measure 
line-of-sight (LOS) stellar velocities from absorption lines to the 
radius of $15^{\prime \prime}$ (6 kpc). 

As we noted above, the surface brightness profile of IC 1541 can be 
decomposed into three exponential segments with different scalelengths. 
By analyzing only the surface photometry, the structure of the galaxy 
can be treated either as a combination of three disks, or of a two-tiered 
disk and a pseudobulge, or of an outer disk and a combined bulge. However, 
stellar dynamics of bulges and disks are different: any bulge is a 
dynamically hot (thick) subsystem, while a disk is a dynamically cool
stellar subsystem, where ordered rotation dominates over random star motions 
(velocity dispersion). Then, we can diagnose the stellar exponential
structures by comparing observed ordered stellar motions with  
predictions of a pure rotation model for a potential with a specified geometry 
and density distribution that is consistent with the observed surface 
brightness distribution. Here, we neglect possible contribution from dark 
matter, but, as is known from observational statistics, the dynamical influence 
of dark matter is negligible in the central regions of massive disk 
galaxies (see, for example, \cite{kassin}).

\begin{figure}
\includegraphics[width=\hsize]{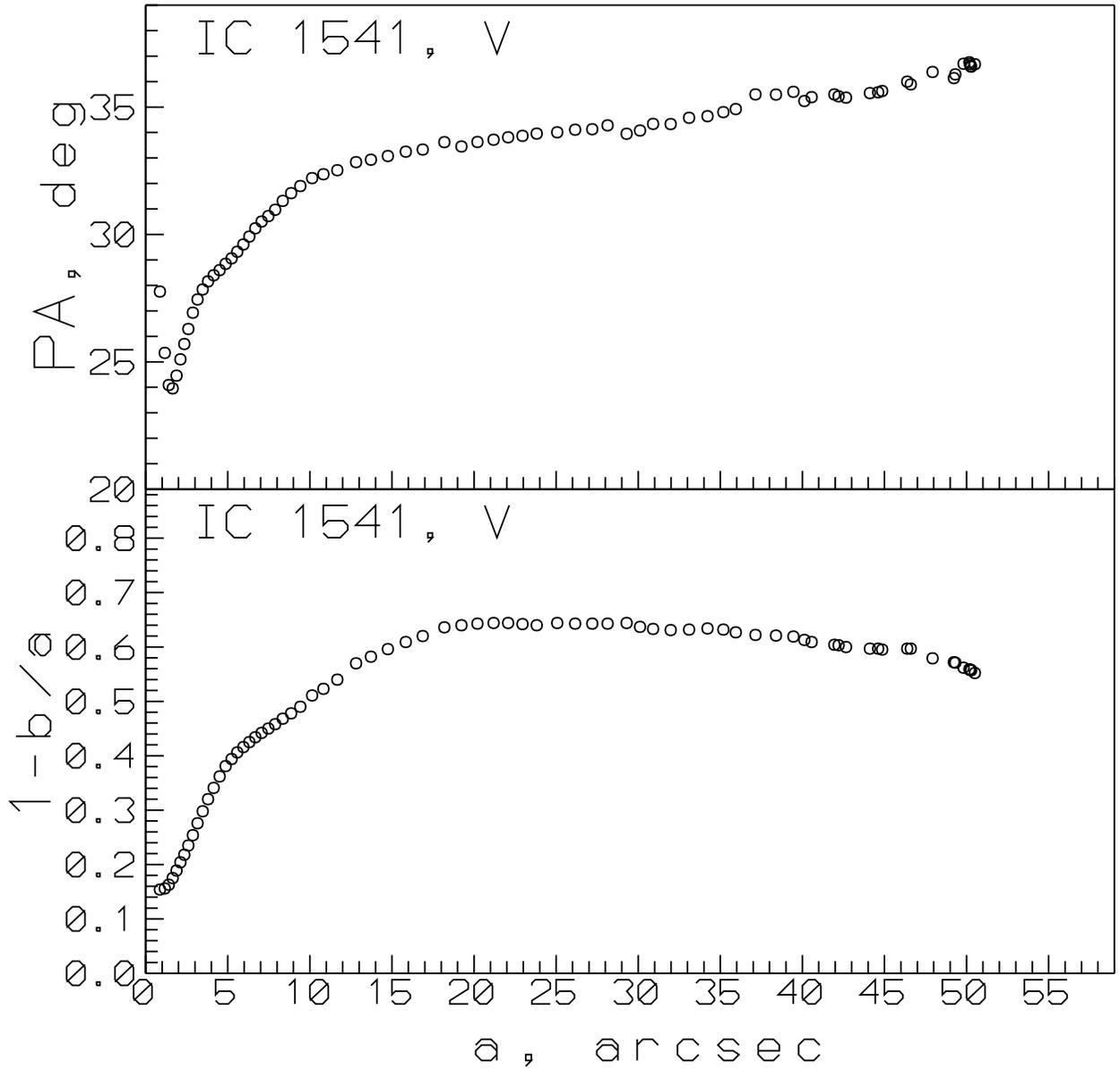}
\caption{The results of isophotal analysis for IC 1541 (in the $V$-band): 
radial dependencies of the major axis position angle and isophote ellipticity.} 
\end{figure}

We have taken the model for the rotation of an axisymmetrical flattened 
stellar system with an exponential profile for the projected surface density 
from \cite{monnetsim}. This paper provides tabulated circular rotation
curves for three values of ratio of the characteristic effective disk 
thickness to the characteristic effective radius (related to the exponential 
scalelength by the factor of 1.678): 0.05, 0.1, and 0.2. The last of these 
three values corresponds to a so-called thick disk, which is typical for
lenticular galaxies \cite{rc3,neistein}, and IC 1541 is a lenticular
galaxy. Figure 5 presents the results of isophotal analysis for IC 1541 
(in the $V$-band). Taking into account the fact that the radial dependence 
of the isophote ellipticity has reached a maximum value (plateau) by the 
radius of $10^{\prime \prime} -15^{\prime \prime}$, i.e. within the zone  
of photometric dominance of the middle exponential component (Fig. 3a), 
the two outer components are supposed to have comparable thicknesses, 
and are most likely disks. The inner exponential component could be 
either a disk or a bulge. Figure 6 compares the observed relative LOS 
velocities for the stellar component of IC 1541 and models from 
\cite{monnetsim} for two possible kinds of the middle disk -- thin
(the ratio of the vertical to the radial scales is 0.05) and thick 
(ratio of the vertical to the radial scales is 0.2); the inner 
exponential component is always taken as a `thick disk'. Based on the 
isophotal analysis of Fig. 5 and the guidelines given in \cite{rc3},
according to which for disks of finite thickness
$$ 
\sin i=\sqrt{\frac{2e-e^2}{1-q_0^2}},
$$
where $e\equiv (1-b/a)$ is the apparent ellipticity of the isophotes
and $q_0$ -- the true ratio of the vertical to the radial scales, we specify
for the `thin' disk an inclination of $67^{\circ}$, while the `thick' disk 
is supposed to be seen edge-on. We took also $M/L_V =3.8$ for the
mass-to-light ratio required to transform the surface brightness profile
into a surface density profile, corresponding to the stellar population
models of \cite{worthey} for a metallicity of $+0.25$ and an age of 5 Gyr.
Figure 6 shows that both models fit fairly the observational data 
for the radius range of $5^{\prime \prime} -15^{\prime \prime}$ 
(the `thick disk' model may be slightly better). Inside the radius of
$5^{\prime \prime}$, where the innermost exponential component dominates 
photometrically, there is a strong discrepance between the model and the 
observations. This is obviously due to the fact that the asymmetric drift 
has not been taken into account (the stellar velocity dispersion at 
the center of IC 1541 reaches 140 km/s), and also possibly because the 
innermost component is not a dynamically cool disk, and its relative thickness 
is much greater than 0.2. As expected, the influence of dark matter is not felt 
at the center of IC 1541. 

Thus, we conclude that the innermost exponential stellar component in IC 1541 is 
a bulge, while the middle component is a disk with the thickness that is typical 
for lenticular galaxies.

\begin{figure}
\includegraphics[width=\hsize]{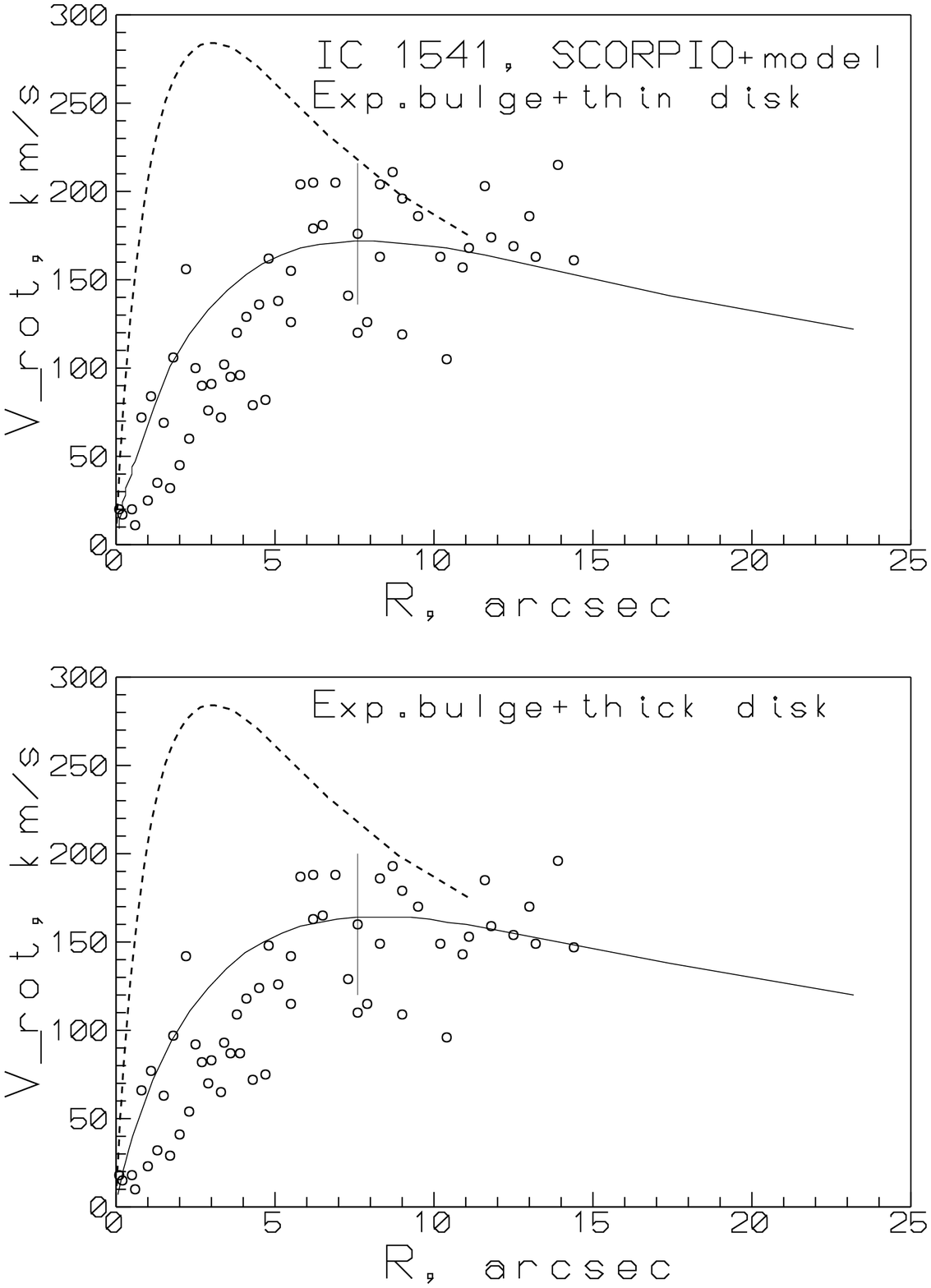}
\caption{Rotation curve for the stellar component of IC 1541. The circles present 
the observational data (the SCORPIO), the solid curve is the dynamical model for the 
inner disk (the middle exponential component with parameters derived from the surface 
photometry, under two assumptions about the disk thickness, see the text for details), 
and the dashed curve shows the dynamical model for the most central exponential 
component assuming that it could be a thick disk. The typical measurement error is 
indicated by the vertical line attached to one of the observational points.}
\end{figure}

\bigskip
\section{MINOR MERGER OF THE LENTICULAR GALAXY NGC 85 }
\medskip

\begin{figure}
\includegraphics[width=\hsize]{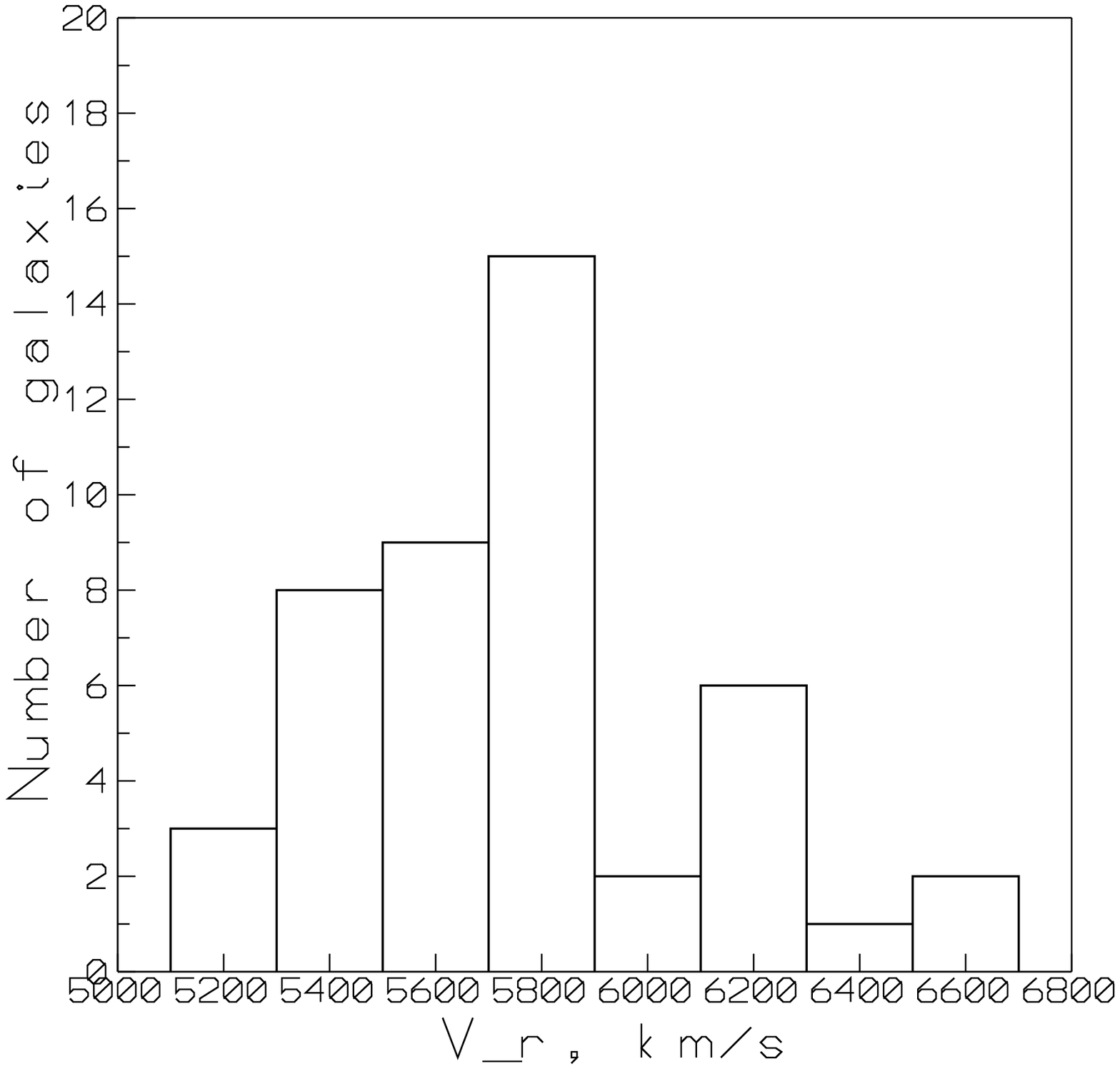}
\caption{Histogram of the distribution of galaxy redshifts in the NGC 80 group 
according to the data of \cite{mahdavi04}. The systemic velocity of the group
and the velocity of the central galaxy, NGC 80, are about 5700 km/s.}
\end{figure} 

\noindent
In our previous paper \cite{we08} where the characteristics of the stellar
populations in the centers of large galaxies of the NGC 80 group were derived, 
we suggested that the subgroup of the giant elliptical galaxy NGC 83 is currently 
accreting onto the NGC 80 group. We were inspired by two facts:
the large difference in the radial velocities of NGC 83 and the dynamical 
center of the group (near to which it is projected), $\sim 500$ km/s, and the
ongoing star formation in the center of this giant elliptical galaxy. Further 
studies whose results we are publishing now have supported our earlier suspicions 
about this. 

Figure 7 shows a histogram of the distribution of galactic redshifts in the area
of the NGC 80 group (taken mainly from \cite{mahdavi04}) for 45 objects within
$\pm 1000$ km/s and within 1.5 Mpc from the center of the group. Although this
distribution can formally be fit with a single Gaussian corresponding to a 
velocity dispersion of 300 km/s, which is normal for massive X-ray galaxy groups, 
there is a concentration of $\sim 10$ objects near the redshift of NGC 83
($v_r \approx 6200$ km/s). Excluding these objects from the NGC 80 group reduces
the estimate of the galaxy velocity dispersion to 224 km/s. The histogram in 
Fig. 7 on its own does not enable us to extract the NGC 83 subgroup in a 
statistically significant way; however, we can note certain common properties 
of the galaxies concentrated near the redshift of NGC 83. First, these are also 
concentrated spatially around NGC 83: most of them are located to the North 
of the group center (NGC 80). Second, almost all of these galaxies display 
ongoing or recent star formation, including NGC 83 itself which is a giant 
elliptical. In addition to NGC 83, SRG 063.023 is a Markarian galaxy (Mrk 1142) 
indicating that its nucleus has ultraviolet excess in the spectrum. The 
giant spiral galaxy MGC $+04 - 02 - 010$ is detected in the Madrid survey of
emission-line galaxies as UCM 0018+2218 \cite{ucm94}, and it is spectrally
classified as a starburst galaxy (SBN) in \cite{ucm96}. Finally, the
lenticular galaxy NGC 85, which is located near NGC 83 in terms of both its 
velocity and its position on the sky plane, and which has been studied by us with 
the MPFS spectrograph, also proves to be unusually young. 

\begin{figure}
\includegraphics[width=\hsize]{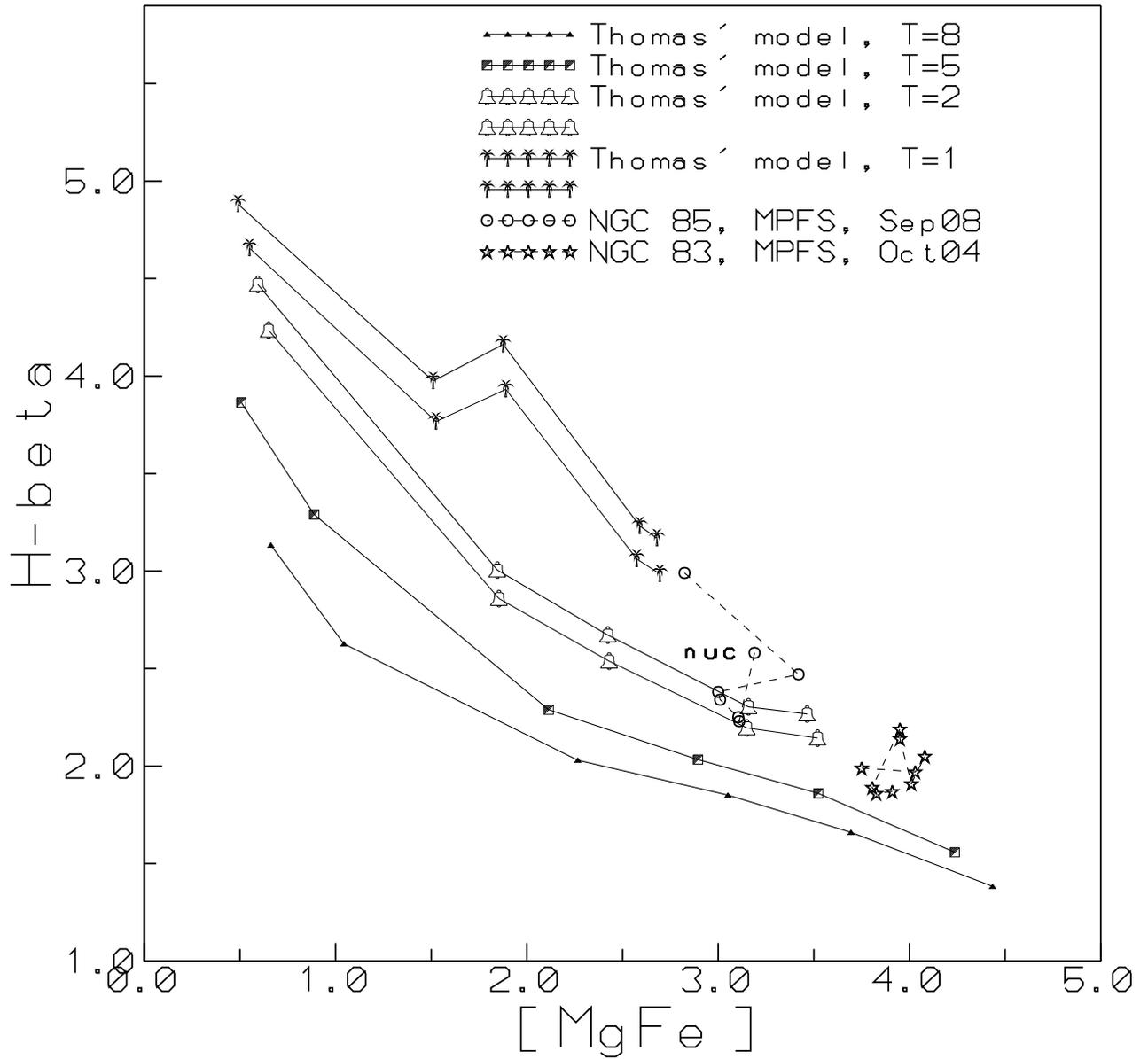}
\caption{Diagnostic `index--index' diagram for the azimuthally-averaged
index measurements in NGC 85 (circles) and NGC 83 (stars).
The measurements are made with the step of $1^{\prime \prime}$ along
the radii, are connected by the dashed lines; the nucleus of NGC 85 is marked
by {\bf nuc}. The model sequences of constant ages from Thomas
et al. (2003) for two values of the ratio [Mg/Fe], 0.0 and $+0.3$,
are also plotted; the total metallicities of the models are $+0.67$,
$+0.35$, 0.00, --0.33, --1.35, --2.25, if one takes the small signs
along the sequences from the right to the left.}
\end{figure}

Figure 8 presents a diagnostic diagram confronting the Lick indices
H$\beta$ and [MgFe]$\equiv \sqrt{ \mbox{Mgb} \langle \mbox{Fe} \rangle }$
for the galaxies NGC 83 and NGC 85 (the data from the MPFS) and models of SSP
by Thomas et al. \cite{tmb03}. This comparison can be used to
separate the effects of age and metallicity, and to determine mean values 
of both these characteristics for a stellar population (weighted with 
star luminosity). The diagram in Fig. 8 demonstrates that, very likely to
NGC 83, NGC 85 has fairly young stars at its center, whose mean ages 
are about 1 to 2 Gyrs in the nucleus and in a ring with the
radius of $5^{\prime \prime} -6^{\prime \prime}$. Obviously, the epochs 
of the star formation bursts in the centers of NGC 85 and NGC 83 are close, 
supporting the idea that these two galaxies share a common fate and are 
both intruders into the NGC 80 group. Opposite to NGC 83, 
NGC 85 contains no nuclear ionized gas, indicating that star formation 
has already ceased in its center. We can identify a likely provoker for 
this star formation burst: a close inspection indicates that the 
galaxy is interacting. The residual brightness distribution (Fig. 1) 
reveals spiral arms, which morphology is consistent with a tidal origin: 
one long and the opposite one short. Which of the neighbors could be 
the perturber? To the east of NGC 85 there is a large spiral galaxy, 
IC 1546, but the difference of their LOS velocities is 300 km/s, 
and such a fast passage would not be able to develop tidal structures. 
A dense, round spheroidal companion is visible to the south of NGC 85; 
could this be the galaxy pulling out the arms from its neighbor? 
Our analysis of the distribution of the $B-V$ color for the residual
brightness field (Fig. 9) disproves this hypothesis: the compact object
to the south of NGC 85 has an anomalously red color to be a nearby galaxy, 
$B-V =1.4$, and is most likely a background galaxy. However, a dense
condensation at the end of the long tidal arm has an appropriate color, 
$B-V =1.1$. This appears to be a half-disrupted satellite of NGC 85, whose
accretion onto the latter has given birth to the observed tidal arms. 
So the lenticular galaxy NGC 85 has recently suffered a minor merger, which has
probably also provoked the nuclear burst of star formation. The bases of the 
tidal arms are very blue, $(B-V)\approx 0.6$, and it is not ruled out
that star formation is ongoing in the ring. If so, this indicates 
that there was quite recently a large amount of gas in the extended disk 
of the galaxy, consistent with the possibility that NGC 85, now a 
lenticular galaxy, was recently a spiral. We conclude that we are 
observing in real time the birth of a lenticular galaxy from a spiral, 
and, although this is occurring within the X-ray halo of the group, the 
mechanism forming the S0 galaxy is, in this case, clearly gravitational and not 
gasdynamical. 

\begin{figure}
\includegraphics[width=\hsize]{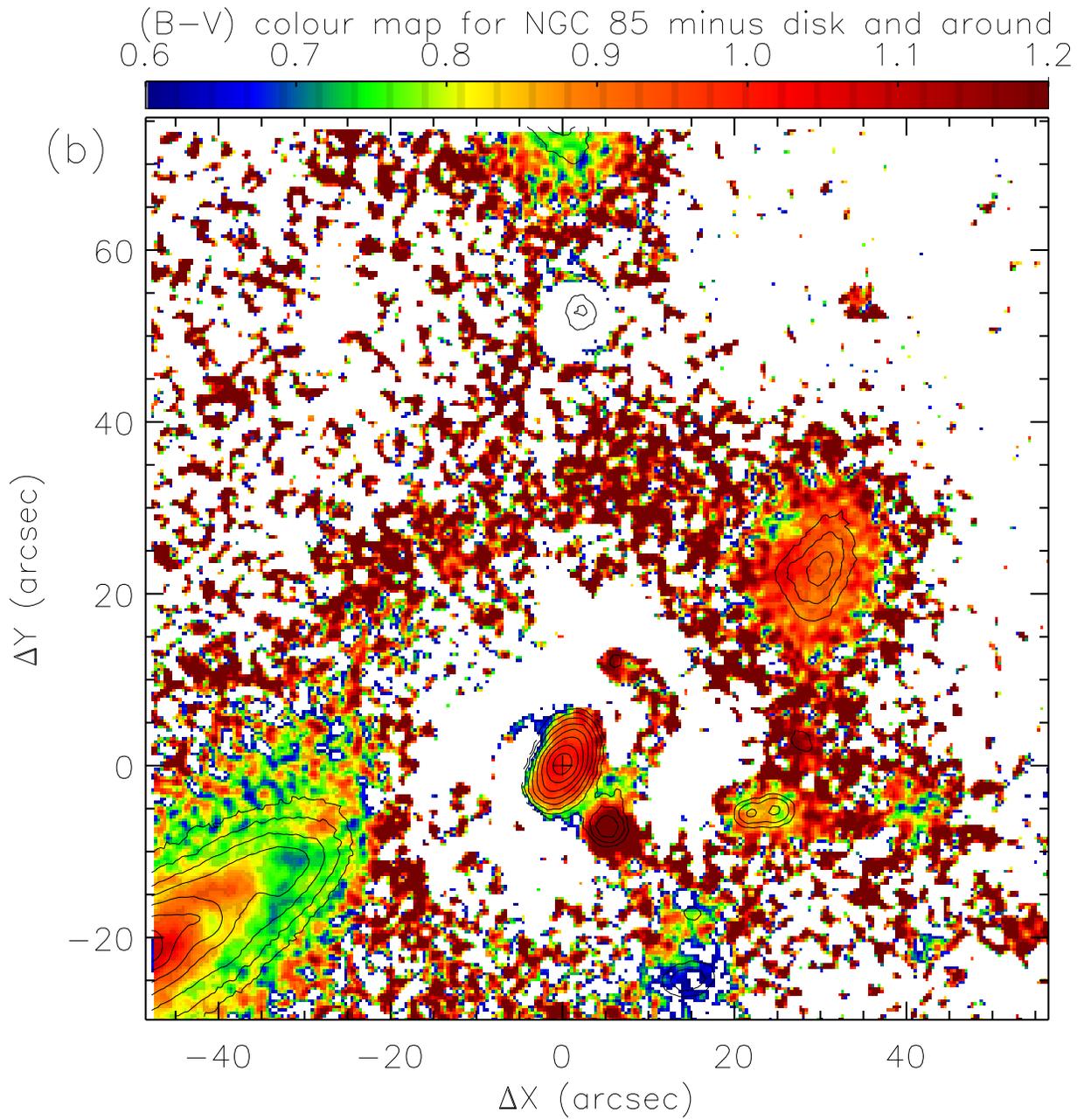}
\caption{Map of $(B-V)$ color for the part of the group around the
galaxy NGC 85 after subtracting the outer disk of NGC 85. The isophotes 
show the residual $V$-surface brightness.}
\end{figure} 

\bigskip
\section{CONCLUSIONS AND DISCUSSION}
\medskip

\noindent
According to the latest views, lenticular galaxies
form from spirals and are shaped completely
in small groups, subsequently infalling with their
groups into clusters where they are currently the dominant
galaxy population (see, for example, \cite{wilman09}).
However, the specific mechanism of this transformation
has not been clear yet: many possible physical mechanisms
are capable on a time scale of only one to four billion years
to remove gas from the disk of a spiral galaxy, to put an end
to star formation, and to heat dynamically the disk, thereby
suppressing spiral structure. These can be divided into
gravitational mechanisms including interactions, harassment,
minor mergers and so no, and gasdynamical mechanisms acting
through the interaction of cool gas of the galactic disk with hot
intergalactic medium. The properties of the NGC 80 group,
which is massive, rich, and possesses a hot X-ray gaseous halo,
make it an ideal place for the formation of lenticular galaxies
through any plausible transformation mechanism. Indeed,
a close inspection of the group reveals a substantial 
number of lenticular galaxies: although rich X-ray
groups are thought to be primarily populated by early-
type galaxies, there are only three elliptical galaxies 
among the several dozen members, and in one of these ellipticals,
NGC 83, we found earlier a massive disk and noticeable current star
formation \cite{we08}. In the current study, we have investigated
the structures of 13 galaxies selected among the most luminous group
members. According to their NED classifications (Table 1),
four of these are lenticulars, and other have later types;
however, a close examination of the fine structure of their
stellar disks suggests that nine are actually lenticulars
(or at least non-spiral).

Our decomposition of the radial surface brightness profiles into
components has shown that most of the lenticular galaxies in the
NGC 80 group have two-tiered, or `anti-truncated' stellar disks
and compact bulges, almost all with exponential brightness profiles
(which are now usually referred to as pseudobulges). This type
of structure is characteristic of both galaxies near the center
of the group, including the central giant S0-galaxy NGC 80, and
galaxies at the periphery, projected at more than 0.5 Mpc
from the group center. We do not see either any systematic
variations in the structure of the disk galaxies from the periphery
to the center of the group, which would have suggested the
transformation of spirals into lenticulars in the course
of their accretion onto the group, nor any morphological
evolution due to sinking into the hot X-ray gaseous halo.
The only environmental effect for which there is evidence is
that four dwarf S0 galaxies ($M_B > -18$) with single-scale,
pure exponential stellar disks are close companions
to the giant galaxies NGC 80 and NGC 86. Thus, 
it appears that the `classical' structure of a lenticular
galaxy, with a single-scale exponential stellar disk, may be
obtained as a consequence of the tidal stripping of 
the outer parts of harassed galaxies.

One of our large lenticular galaxies, NGC 85,
shows signatures of recent stimulated nuclear star formation:
the mean age of the stars in its center is about one or two Gyr,
and subtracting a model for the outer stellar
disk reveals the presence of spiral arms with blue 
bases that are clearly tidal in origin. It is possible
that we are observing here the recent transformation of
a spiral into a lenticular galaxy, accompanied by the
star formation burst in the galactic center. NGC 85 is
located not far from the center of the group, but, 
according to its radial velocity, it is associated with 
NGC 83, and apparently has joined the NGC 80 group
only recently, together with NGC 83. The remnants 
of another, small galaxy can be seen at the
end of a long tidal arm giving evidence for a minor
merging. All this supports the idea that the mechanism
for the transformation of spirals into lenticulars could
be gravitational, and in particular, minor mergers would be
especially effective. The properties of NGC 85 and the
presence of well-formed lenticular galaxies such as
IC 1541 \cite{we08}, IC 1548 \cite{we08}, and NGC 94 \cite{burphot}
at the group periphery, outside the X-ray halo, seem to be
strong arguments for the dominant mechanisms responsible
for the transformation of spirals into lenticulars in groups
to be gravitational ones, rather than gasdynamical interaction
with hot, intergalactic gas.

\bigskip
\section{ACKNOWLEDGMENTS}
\medskip 

\noindent
The data analyzed here were obtained on the 6-m 
telescope of the Special Astrophysical Observatory 
of the Russian Academy of Sciences, which is supported
by the Ministry of Education and Science of 
the Russian Federation (registration number 01-43). 
We thank A.A. Smirnova for supervising the MPFS 
observations. Our analysis made use of the Lyon-Meudon
Extragalactic Database (LEDA), which is
maintained by the LEDA team at the Lyon Observatory CRAL
(France), and the NASA/IPAC Extragalactic Database (NED),
which is managed by the Jet Propulsion Laboratory of the
California Institute of Technology by contract to the
NASA (USA). This work was supported by the Russian Foundation
for Basic Research (project no. 07-02-00229a).


\begin{thebibliography}{}

\bibitem[1]{hubble}
Hubble E. "Realm of the Nebula". New Haven: Yale Univ. Press (1936)

\bibitem[2]{fasano00}
Fasano G., Poggianti B.M., Couch W.J., et al. Astrophys. J. {\bf  542}, 673 (2000)

\bibitem[3]{wilman09}
Wilman D.J., Oemler A., Jr., Mulchaey J.S., et al. Astrophys. J. {\bf 692}, 298 (2009)

\bibitem[4]{burstein}
Burstein D. Astrophys. J. {\bf 234}, 435 (1979)

\bibitem[5]{mh01}
M$\ddot o$llenhoff C., Heidt J. Astron. Astrophys. {\bf 368}, 16 (2001)

\bibitem[6]{quilis}
Quilis V., Moore B., Bower R. Science {\bf 288}, 1617 (2000)

\bibitem[7]{zasov}
Zasov A.V. Pis'ma v AZh {\bf 4}, 487 (1978)

\bibitem[8]{byrdvalt}
Byrd G., Valtonen M. Astrophys. J. {\bf 350}, 89 (1990)

\bibitem[9]{moore96}
Moore B., Katz N., Lake G., et al.  Nature {\bf 379}, 613 (1996)

\bibitem[10]{moore99}
Moore B., Lake G., Quinn T., Stadel J. MNRAS {\bf 304}, 465 (1999)

\bibitem[11]{oemler}
Oemler A. Astrophys. J. {\bf 194}, 1 (1974)

\bibitem[12]{dressler}
Dressler A. Astrophys. J. {\bf 236}, 351 (1980)

\bibitem[13]{ButchOem}
Butcher H.R. \& Oemler A. Astrophys. J. {\bf 226}, 559 (1978)

\bibitem[14]{gh_cat}
Geller M.J., Huchra J.P. Astrophys. J. Suppl. Ser. {\bf 52}, 61 (1983)

\bibitem[15]{mahdavi00}
Mahdavi A., Bohringer H., Geller M.J., Ramella M. Astrophys. J.
{\bf 534}, 114 (2000)

\bibitem[16]{ramella_cat}
Ramella M., Geller M.J., Pisani A., da Costa L.N. Astron. J. {\bf 123},
2976 (2002)

\bibitem[17]{wbl99}
White R.A., Bliton M., Bhavsar S.P., et al. Astron. J. {\bf 118},
2014 (1999)

\bibitem[18]{mahdavi04}
Mahdavi A., Geller M.J. Astrophys. J. {\bf 607}, 202 (2004)

\bibitem[19]{we08}
Sil'chenko O.K., Afanasiev V.L. Astronomy Reports {\bf 52}, 875 (2008)

\bibitem[20]{scorpio}
Afanasiev V.L., Moiseev A.V. Astron. Letters {\bf 31}, 194 (2005)]

\bibitem[21]{poulain}
Poulain P. Astron. Astrophys. Suppl. Ser. {\bf 72}, 215 (1988)

\bibitem[22]{mpfs}
Afanasiev V.L., Dodonov S.N., Moiseev A.V.// Proc. of the Conf.
   "Stellar dynamics: from classic to modern", St. Petersburg, 2001/
    Eds. Osipkov L.P. and Nikiforov I.I., Saint Petersburg Univ. press, p.103

\bibitem[23]{woretal94}
Worthey G., Faber S.M., Gonz\`alez J.J., Burstein D.
Astrophys. J. Suppl. Ser. {\bf 94}, 687 (1994)

\bibitem[24]{tmb03}
Thomas D., Maraston C., Bender R. MNRAS {\bf 339}, 897 (2003)

\bibitem[25]{moiseev}
Moiseev A.V., Vald{\'e}s J.R., Chavushyan V.H. Astron. Astrophys., {\bf 421}, 433 (2004)

\bibitem[26]{sersic}
S\'ersic J.L. Atlas de Galaxies Australes. Cordoba: Observatorio
Astronomico (1969)

\bibitem[27]{freeman}
Freeman K.C. Astrophys. J. {\bf 160}, 767 (1970)

\bibitem[28]{dejong}
de Jong R.S. Astron. Astrophys. {\bf 313}, 45 (1996)

\bibitem[29]{iodice}
Iodice E., D'Onofrio M., Capaccioli M. Astrophys. and Space Science
{\bf 276}, 869 (2001)

\bibitem[30]{graham}
Graham A.W., de Blok W.J.G. Astrophys. J. {\bf 556}, 177 (2001)

\bibitem[31]{donofrio}
D'Onofrio M. MNRAS {\bf 326}, 1517 (2001)

\bibitem[32]{bureau06}
Bureau M., Aronica G., Athanassoula E., et al. MNRAS {\bf 370}, 753 (2006)

\bibitem[33]{we09}
Sil'chenko O.K., Moiseev A.V., Afanasiev V.L. Astrophys. J. {\bf 694}, 1550 (2009)

\bibitem[34]{ucm96}
Gallego J., Zamorano J., Rego M., et al. Astron. Astrophys. Suppl.
Ser. {\bf 120}, 323 (1996)

\bibitem[35]{we03}
Sil'chenko O.K., Koposov S.E., Vlasyuk V.V., Spiridonova O.I. Astronomy
Reports {\bf 47}, 88 (2003)

\bibitem[36]{kassin}
Kassin S.A., de Jong R.S., Weiner B.J. Astrophys. J. {\bf 643}, 804 (2006)

\bibitem[37]{monnetsim}
Monnet G., Simien F. Astron. Astrophys. {\bf 56}, 173 (1977)

\bibitem[38]{rc3}
de Vaucouleurs G., de Vaucouleurs A. Corwin H.G., Jr., et al.
"Third Reference Catalogue of Bright Galaxies. Volume I: Explanations and References".
New York: Springer (1991)

\bibitem[39]{neistein}
Neistein E., Maoz D., Rix H.-W., Tonry J.L. Astron. J. {\bf 117},
2666 (1999)

\bibitem[40]{worthey}
Worthey G. Astrophys. J. Suppl. Ser. {\bf 95}, 107 (1994)

\bibitem[41]{ucm94}
Zamorano J., Rego M., Gallego J., et al. Astrophys. J. Suppl. Ser.
{\bf 95}, 387 (1994)

\bibitem[42]{burphot}
Kalloghlian A.T., Nikoghossian E.H. Astrophysics {\bf 36}, 193 (1994)


\end{thebibliography}
\end{document}